\documentclass[%
 reprint,
 amsmath,amssymb,
 aps,
]{revtex4-2}

\usepackage{cases}
\usepackage{graphicx}
\usepackage{CJK}
\usepackage{subfigure}
\usepackage{natbib}
\usepackage{xcolor}
\usepackage{tabularx}
\usepackage{bm}
\usepackage{threeparttable}
\usepackage{dcolumn}
\usepackage{lipsum}
\usepackage{aas_macros}

\begin{document}

\newcommand{\hanyue}[1]{\textbf{\color{orange}Hanyue: #1}}
\newcommand{\daniel}[1]{\textbf{\color{purple}Daniel: #1}}

\preprint{APS/123-QED}

\title{A Spectroscopic Search for Optical Emission Lines from Dark Matter Decay}

\author{Hanyue Wang$^{1,2}$}
\author{Daniel~J.~Eisenstein$^{1,2}$}
\author{Jessica Nicole Aguilar$^{3}$}
\author{Steven Ahlen$^{4}$}
\author{Stephen Bailey$^{3}$}
\author{David~Brooks$^{5}$}
\author{Todd~Claybaugh$^{3}$}
\author{Axel~de la Macorra$^{6}$}
\author{Peter~Doel$^{5}$}
\author{Jaime~E.~Forero-Romero$^{7}$}
\author{Anthony Kremin$^{3}$}
\author{Michael~E.~Levi$^{3}$}
\author{Marc~Manera$^{8,9}$}
\author{Ramon~Miquel$^{9,10}$}
\author{Claire~Poppett$^{3}$}
\author{Mehdi Rezaie$^{11}$}
\author{Graziano~Rossi$^{12}$}
\author{Eusebio~Sanchez$^{13}$}
\author{Michael~Schubnell$^{14}$}
\author{Gregory~Tarl\'{e}$^{14}$}
\author{Benjamin A.~Weaver$^{15}$}
\author{Zhimin Zhou$^{16}$}

\affiliation{$^{1}$Department of Astronomy, Harvard University, 60 Garden St., Cambridge MA 02138, USA}
\affiliation{$^{2}$ Center for Astrophysics $|$ Harvard \& Smithsonian, 60 Garden St., Cambridge, MA 02138, USA}
\affiliation{$^{3}$Lawrence Berkeley National Laboratory, 1 Cyclotron Road, Berkeley, CA 94720, USA}
\affiliation{$^{4}$Physics Dept., Boston University, 590 Commonwealth Avenue, Boston, MA 02215, USA}
\affiliation{$^{5}$Department of Physics \& Astronomy, University College London, Gower Street, London, WC1E 6BT, UK}
\affiliation{$^{6}$Instituto de F\'{\i}sica, Universidad Nacional Aut\'{o}noma de M\'{e}xico,  Cd. de M\'{e}xico  C.P. 04510,  M\'{e}xico}
\affiliation{$^{7}$Departamento de F\'isica, Universidad de los Andes, Cra. 1 No. 18A-10, Edificio Ip, CP 111711, Bogot\'a, Colombia}
\affiliation{$^{8}$Departament de F\'{i}sica, Serra H\'{u}nter, Universitat Aut\`{o}noma de Barcelona, 08193 Bellaterra (Barcelona), Spain}
\affiliation{$^{9}$Institut de F\'{i}sica d’Altes Energies (IFAE), The Barcelona Institute of Science and Technology, Campus UAB, 08193 Bellaterra Barcelona, Spain}
\affiliation{$^{10}$Instituci\'{o} Catalana de Recerca i Estudis Avan\c{c}ats, Passeig de Llu\'{\i}s Companys, 23, 08010 Barcelona, Spain}
\affiliation{$^{11}$Department of Physics, Kansas State University, 116 Cardwell Hall, Manhattan, KS 66506, USA}
\affiliation{$^{12}$Department of Physics and Astronomy, Sejong University, Seoul, 143-747, Korea}
\affiliation{$^{13}$CIEMAT, Avenida Complutense 40, E-28040 Madrid, Spain}
\affiliation{$^{14}$Department of Physics, University of Michigan, Ann Arbor, MI 48109, USA}
\affiliation{$^{15}$NSF's NOIRLab, 950 N. Cherry Ave., Tucson, AZ 85719, USA}
\affiliation{$^{16}$National Astronomical Observatories, Chinese Academy of Sciences, A20 Datun Rd., Chaoyang District, Beijing, 100012, P.R. China}

\begin{abstract}
    We search for narrow-line optical emission from dark matter decay by stacking dark-sky spectra from the Dark Energy Spectroscopic Instrument (DESI) at the redshift of nearby galaxies from DESI's Bright Galaxy and Luminous Red Galaxy samples. Our search uses regions separated by 5 to 20 arcsecond from the centers of the galaxies, corresponding to an impact parameter of approximately $50\,\rm kpc$.  
    No unidentified spectral line shows up in the search, and we place a line flux limit of $10^{-19}\,\rm{ergs}/\rm{s}/\rm{cm}^{2}/\rm{arcsec}^{2}$ on emissions in the wavelength range of $2000$ -- $9000 \,\mathring{\rm A}$. 
    This places the tightest constraints yet on the two-photon decay of dark matter in the mass range of 5 to $12\,\rm eV$, with a particle lifetime exceeding $3\times 10^{25}\,\rm s$.
    This detection limit also implies that the line surface brightness contributed from all dark matter along the line of sight is at least two orders of magnitude lower than the measured extragalactic background light (EBL), ruling out the possibility that narrow optical-line emission from dark matter decay is a major source of the EBL. 
\end{abstract}

\maketitle

\section{Introduction} \label{sec:intro}

Dark matter is an important cornerstone for the standard $\Lambda$CDM cosmological model and a main component of current large-scale structure formation history. Cosmological observations have shown that a comparable amount of dark matter is needed at low redshift for galaxy rotation curves, strong gravitational events, and galaxy clustering. Although dark matter is often considered as stable long-lived particles, deviations from this empirical limit would provide critical information about the true nature of dark matter.
There are many hypotheses about how particle candidates beyond the Standard Model appear from dark matter decay, for instance, sterile neutrinos \citep{1994PhRvL..72...17D}, gravitinos \citep{1995hep.ph....3210M} and axions \citep{1999PhRvL..82.4180C}. These models typically describe how dark matter particles are able to decay through weak coupling with Standard Model particles to radiation, such as two photons \citep{photons1, photons2} or one photon and one other particle (such as Z boson or neutrino) \citep{photon+neutrino}. 
In certain scenarios, weak coupling with baryons prompts the decay of dark matter into a photon line. One such example is the decay of axion-like particles, which are known to be unstable and can decay into two photons \citep{axion,axion2,axion3}. Additionally, in the neutrino minimal standard model ($\nu$MSM), decay of sterile neutrinos as dark matter candidates emit monoenergetic photons \citep{neutrino}.  

Dark matter decay properties can be constrained with different processes. \cite{1988PhRvL..61..510C, 2022PTEP.2022a3F01M} study the decay behavior directly from a laboratory setup or detectors and \cite{2021JCAP...10..040H} used cosmological N-body simulations to make predictions. There are also ongoing endeavors to probe these properties with high-energy astronomical observations at the galactic center and dwarf galaxies \citep{dwarf}, of the gamma-ray extragalactic background \citep{gamma-ray}, and in the X-ray \citep{x-ray}. 
\cite{3.5keV} discovered an unidentified X-Ray emission at around $3.5\,\rm keV$ in the stacked spectrum of 73 galaxy clusters, and one hypothesis to explain this new line is dark matter decay, especially the decay of sterile neutrinos in keV mass range. Although blank sky observations from XMM-Newton have declined this hypothesis \citep{inconsistent}, our work is motivated by a similar idea of monoenergetic dark matter decay into a single photon line. Similar searches have been done with the MUSE spectroscopic observations of the dwarf galaxies \citep{MUSE1, MUSE2}, and \cite{VIMOS} looked for optical line emission from decaying relic axions with spectra of galaxy clusters from VIMOS (Visible Multi-Object
Spectrograph) at the Very Large Telescope.
 
In this work, we use a high-fidelity map of the location of galaxies/dark matter to cross-correlate with the spectral map of the dark sky or faint background object observations to look for the additional spectral line that corresponds with monoenergetic dark matter decay in the optical range, based on a large dataset from Dark Energy Spectroscopic Instrumentation (DESI) \citep{DESI2016a, DESI2016b, DESI2022, focal_plane, corrector}. 
As dark matter halos are extended much further compared with optically observable regions of galaxies, we target the far outskirts of the bright galaxies at redshift around $0.2$ and of luminous red galaxies at redshift around $0.7$ to avoid the influence of optical light.  Compared with annihilation which follows the square of density profile $\rho^{2}$, dark matter decay scales by $\rho$ \citep{DMdensity}, which makes far outskirts of galaxies still a reasonable place to search for evidence. We stack the spectra in the rest-frame of corresponding nearby galaxies to avoid observational or instrumental systematics that could arise at fixed observed wavelength. 
Although individual sources of dark matter are not detectable, we search for emission by co-adding large number of spectra with accumulations of long exposure times. 

This paper is structured as follows. In Section~\ref{sec: data} we describe the data samples being used together with selection criteria and preparations. In Section~\ref{sec: correlation} we present our methods for coaddition to obtain the stacked spectrum. We show our results in Section~\ref{sec:result} about line detection and the detection limit, and we compute the mass-luminosity relation for dark matter decay in Section~\ref{sec: comp}. Finally, we summarize and discuss the results further in Section~\ref{sec:dis}.

\section{Data} \label{sec: data}

This study makes use of data from the DESI survey, including the foreground galaxy position and redshift data and the spectra from both sky fibers and faint background galaxies.  We use the data from the Year 1 data set, collected through June 2022, with processing via the ``iron'' version of the data compilation \citep{iron_pipeline}. It will be released as part of DESI Data Release 1 (DR1), in a manner similar to that of the Early Data Release \citep{SV_KP, EDR_KP}

DESI Target selection is based on the Legacy Imaging Surveys Data Release 9 \citep{LS, BASS}, specifically the $g, r, z$ optical bands and the the Wide-field Infrared Survey Explorer(\textit{WISE}, \citep{WISE}) photometry.
We use the bright galaxies (BGs) at low redshift from the Bright Galaxy Survey (BGS) \citep{BGS_final} and only select those that the redshift pipeline assesses to have bitmask $\tt ZWARN=0$, as these have been found to yield a robust set with few redshift errors. We choose 5914291 galaxies from the BGS in total with an average redshift of $0.24$. 
To enlarge the wavelength range of our search, massive galaxies at higher redshift are also selected from DESI Luminous Red Galaxy (LRG) Sample in the redshift range $0.3 < z \lesssim 1.0$ \citep{LRG_new} . We make use of 2492017 LRGs in total with an average redshift of 0.72.

We then collect the sample of spectra from the dark-time survey tiles that we will consider for stacking. The dark-time survey has longer exposures and lower moon contamination, so it is better for searching for faint signals.  We seek spectra that have relatively little light from their intended target.  Many such spectra exist because of the sky fibers, intentionally placed on areas that were blank in the optical imaging used for target selection and spectral calibration \citep{sky}. These sky fibers point at pixel-level blank sky locations that are expected to include minimal flux from astronomical sources when integrated over a DESI fiber. But many DESI targets are also rather faint, and we opt to use these as well, provided that they are at a well-separated redshift.
To keep the spectra in a faint enough level, the faint background galaxies are selected within a fiber total flux range of $F_{G}\leq0.63$, $F_{R}\leq1.0$, and $F_{Z}\leq1.58$ in unit of nanomaggies in the corresponding photometric bands $g\,(3980-5480\,\mathring{\rm A}), r\,(5680-7160\,\mathring{\rm A}), z\, (7100-8570\,\mathring{\rm A})$. We do not use quasar targets, as these are often brighter and have more variations in their spectra. 
To avoid including spectral pixels of known emission from these spectra of faint background targets, we further pre-process these spectra by masking wavelengths corresponding to strong oxygen and hydrogen emission lines at the redshift of the target. This is simply to avoid including pixels with large narrow-band lines that we know have a familiar physical cause, but that will add noise to the stack.  

We convert the angular positions of the primary galaxy samples into Cartesian coordinates on a unit sphere and 
search for nearby sky or faint background objects as targets for the spectral map. We set a separation threshold of $20''$ to look for pairing within this angular distance range. 
When pairing with the faint background targets, we reject the pairs with similar redshift,  $\left\|z_{\rm tgt}-z_{\rm bgs}\right\|<0.05\times(1+z_{\rm bgs})$, to sharply reduce any correlations with the primary object.

With this search, we find 80472 sky fiber spectra around the bright galaxies and 40659 sky fiber spectra around LRGs. We also find 437502 background galaxies fibers on faint background objects around BGs and 150958 background galaxies fibers around LRGs. For each foreground galaxy-faint spectrum pair, the spectra of the sky (or qualified faint galaxy targets) are shifted to the rest-frame wavelength of the corresponding primary galaxy. 
We note that the DESI errors on redshift calibration are typically better than $9\,\rm km/s$ for the BGS and $42\,\rm km/s$ for the LRGs \citep{SV_KP}, which are much smaller than the $200\,\rm km/s$ line width we are integrating over for coaddition which is described in Section~\ref{sec: correlation}. Further, the instrumental line-broadening effect is negligible with a spectral resolution of approximately $40\,\rm km/s$ \citep{DESI2022}, which is again much smaller than the line width we are integrating over for coaddition.

The fiber positioners can block neighbouring fibres from targeting certain objects, and objects can be missed in regions with high number density of targets. These fibre collision effects will suppress the number of pairs with small angular separation, but because DESI returns to one sky region multiple times with different targets, there are plenty of small-angle measurements and when averaging within the angular bin, the fiber collision effect will drop out.

\section{Coaddition} \label{sec: correlation}

To stack the faint spectra in the rest frame of nearby galaxies, we co-add the rest-frame spectra into a wavelength grid from $2000\,\mathring{\rm A}$ to $10000\,\mathring{\rm A}$ with $1\,\mathring{\rm A}$ bins. Coaddition is done by assigning individual pixels from each spectrum into single bins of the final wavelength grid.
The final stacked spectrum $\xi$ after co-addition is given by 
\begin{equation}
    \xi = \frac{1}{\sum_{i=1}^{N}\omega_{i}}\sum_{i=1}^{N}\omega_{i}F_{i},
\end{equation}
where $N$ is the total number of pixels in each bin of the grid, $\omega_{i}=1/\sigma_{i}^{2}$ is the inverse variance of flux at each pixel, and $F_{i}$ is the flux.  Masked pixels are given the inverse variance of zero.

First, we co-added all available spectra within the impact parameter of $50\,\rm kpc$ to create an overall stacked spectrum. Moreover, to investigate the relation between distance and luminosity profile, we classify the faint spectra based on their angular separation or physical separation from the corresponding primary galaxies and stacked the spectra in these different bins. The angular separation bins are logarithmically spaced from $2.5$ to $20$ arcsecond. The physical separation is calculated using a flat cosmological model with matter density $\Omega_{m}=0.3$ and cosmological constant density $\Omega_{\Lambda}=0.7$ at the redshift of the foreground galaxy, and then we use four logarithmically spaced bins of separation from $10$ to $160\,\rm kpc$.


We search for prominent lines in these stacked spectra. For each stack, we utilize a Savitzky–Golay filter with a window size of $205\,\mathring{\rm A}$ to smooth the data and then remove the smoothing curve to rescale the average flux level around zero. Then, we convolve the spectrum with a 1D Gaussian kernel to calculate the line flux. We use two such kernels with different sizes. The primary narrow kernel with a standard deviation of $3\,\mathring{\rm A}$ is physically motivated by the general line broadening of an intrinsically narrow line due to the normal velocity dispersion of halos of dark matter particles. However, the line could also be intrinsically broader. To investigate this possibility, we also make use of a broad kernel with a standard deviation of $15\,\mathring{\rm A}$. This kernel size is limited by the fact that the high-pass filtering needed to remove the continuum flux left over from the faint galaxies would artificially remove signals of broader lines.

\begin{figure}[htb]
    \centering
    \includegraphics[width=0.98\columnwidth]{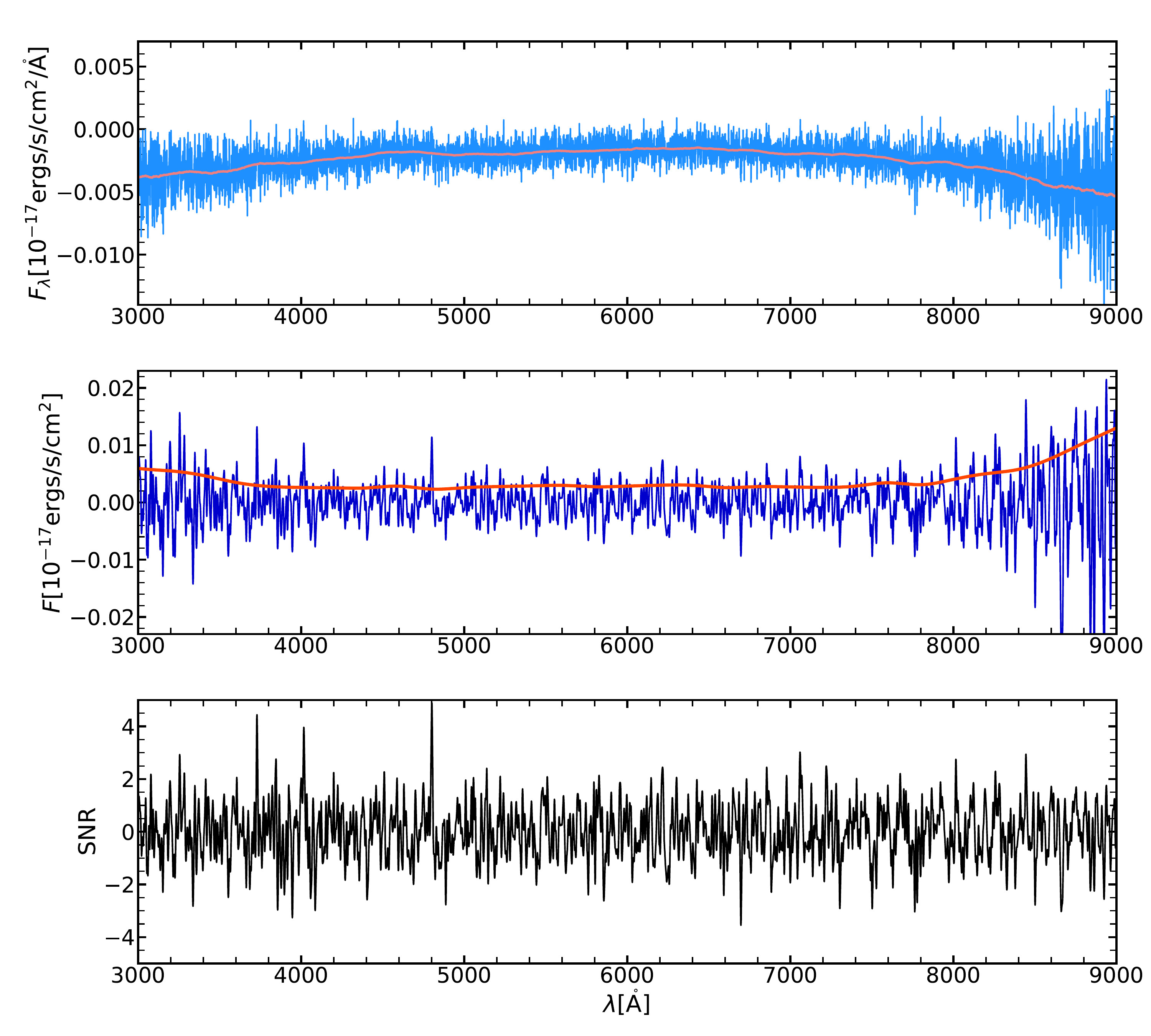}
    \caption{An example of data processing for pairs of bright galaxies and sky fibers. The upper panel shows the original stacked spectrum after coaddition in blue, and the pink curve is the smoothed curve generated by the Savitzky–Golay filter. The middle panel (blue) shows the Gaussian convolution with a $3\,\mathring{\rm A}$ kernel of the spectrum after subtracting the smooth continuum to find out the line flux. The red line is the estimated error curve from calculating half the interval between the $16\%$ and $84\%$ quantiles in successive subsets of adjacent data points. The last panel shows the signal to noise ratio.}
    \label{fig:eg}
\end{figure}

The final step is to find the signal-to-noise ratio (SNR) at each wavelength. We first calculate the robust estimated error using half the interval between the $16\%$ and $84\%$ quantiles in subsets of data within $400\,\mathring{\rm A}$ bins separated by $200\,\mathring{\rm A}$, and then fit a smooth curve with these points. The SNR is generated by the convolution spectrum over the error level. An example of this process is displayed in Figure \ref{fig:eg}, which shows the stacked spectrum for all sky fiber-bright galaxy pairs.

We constrain the wavelength range from $3000\,\mathring{\rm A}$ to $9000\,\mathring{\rm A}$ for BG pairs as most of the pairs could inform these wavelengths. For pairs with LRGs, we switch the wavelength range to $2000$--$7000\,\mathring{\rm A}$ to take advantage of the higher redshift.

\section{Results} \label{sec:result}

\begin{figure}[htb]
    \centering
    \includegraphics[width=0.98\columnwidth]{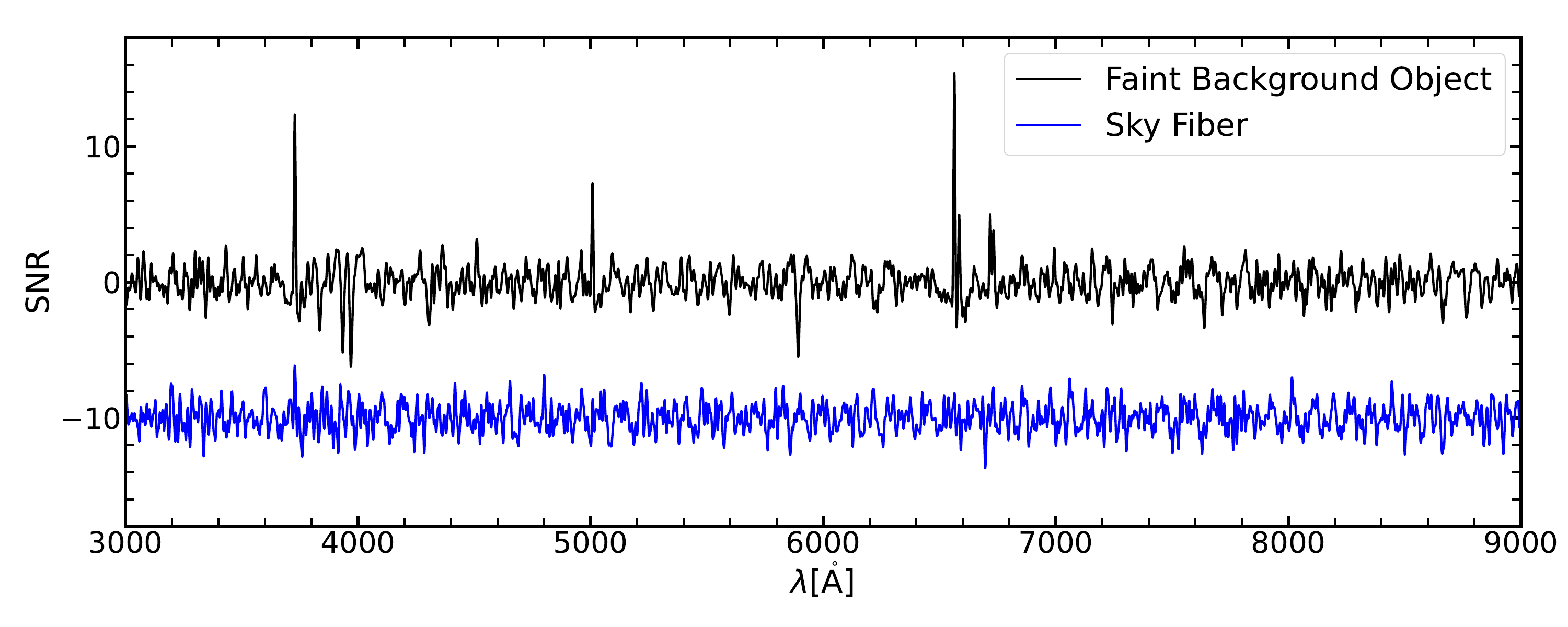}
    \caption{The SNR of all pairs of BGs and faint background objects in black and SNR of all sky pairs in blue. The SNR of sky pairs is shifted by $-10$ from $0$ to $-10$ on the plot. }
    \label{fig:snr}
\end{figure}

\begin{figure}[htb]
    \centering
    \includegraphics[width=0.98\columnwidth]{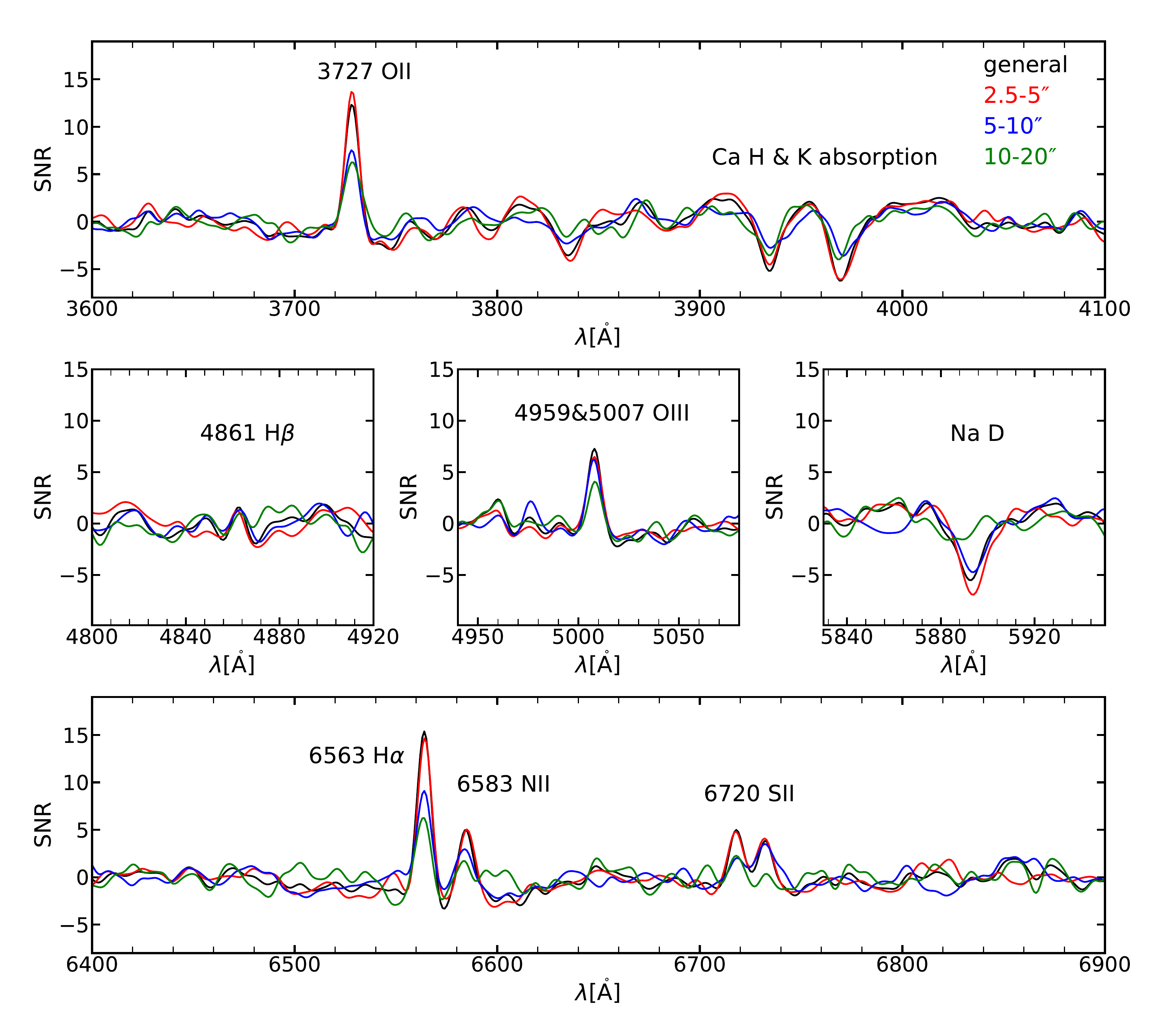}
    \includegraphics[width=0.98\columnwidth]{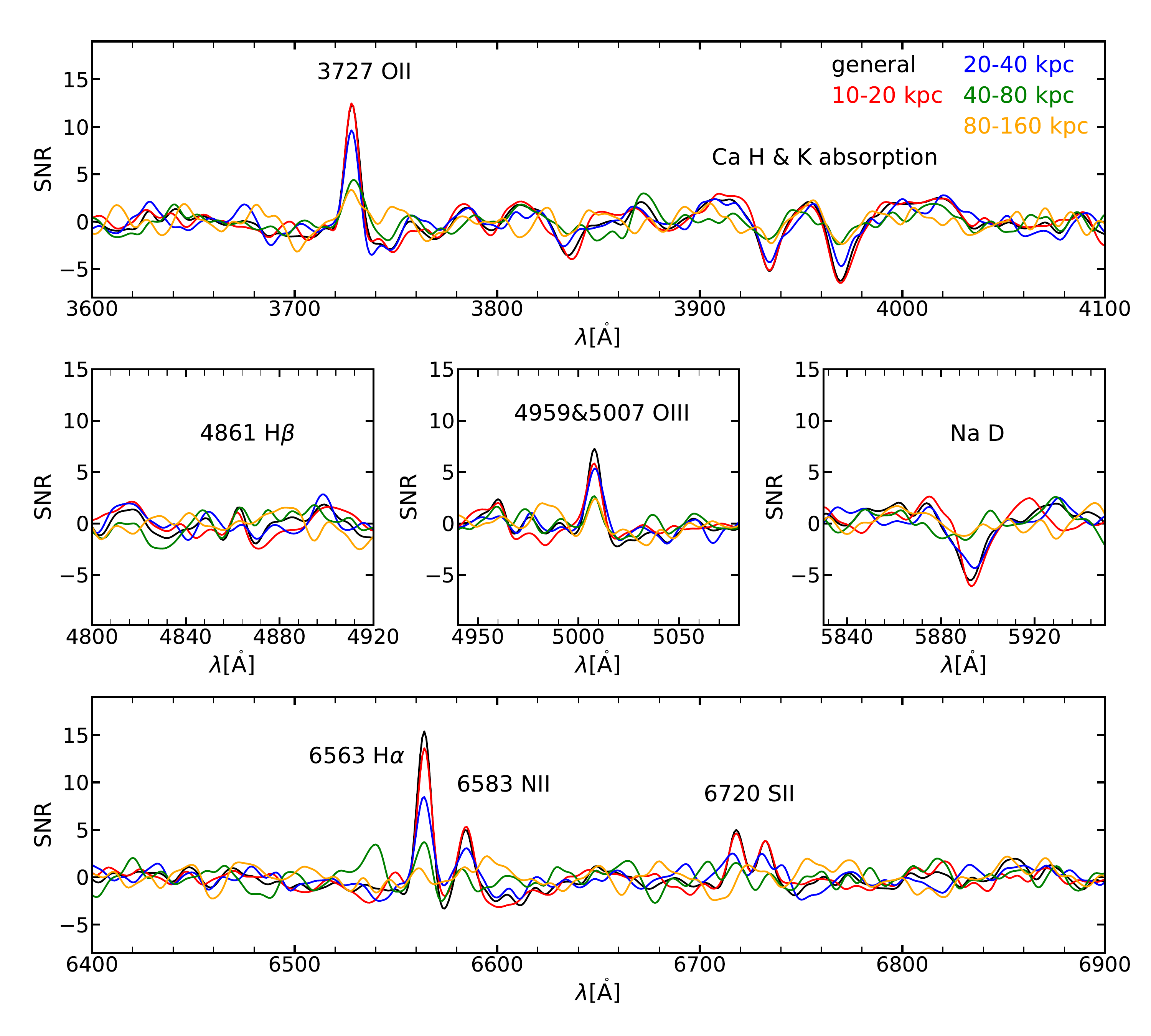}
    \caption{The SNR of all available bright galaxy-faint background object pairs zoomed in around prominent lines. The upper panel shows the result in angular separation bins, and the lower panel shows the result in physical separation bins. The bins are specified on the upper-right corner of each panel. The black ``General" curve is the sum of all available pairs with a separation below the threshold of $20''$. }
    \label{fig:snr_bins}
\end{figure}

We first focus on the bright galaxies at lower redshift. We show the SNR of all bright galaxy-sky fiber pairs and bright galaxy-faint background target pairs in Figure \ref{fig:snr}.
We did not find unknown signals that might be correlated with dark matter decay in the galaxy halo, but we do recognize several famous spectral lines in the SNR of the cross-correlation between faint targets and bright galaxies, including the [O\uppercase\expandafter{\romannumeral2}] emission doublet at $3726$ and $3729\,\mathring{\rm A}$, Calcium H and K absorption at $3969$ and $3934\,\mathring{\rm A}$, [O\uppercase\expandafter{\romannumeral3}] lines at $4959$ and $5007\,\mathring{\rm A}$, Sodium D absorption at $5890$ and $5896\,\mathring{\rm A}$, H$\rm \alpha$ at $6563\,\mathring{\rm A}$, [N\uppercase\expandafter{\romannumeral2}] at $6583\,\mathring{\rm A}$, and two lines from singly-ionized sulfur ([S\uppercase\expandafter{\romannumeral2}]) at $6716$ and $6731\,\mathring{\rm A}$. Figure \ref{fig:snr_bins} shows a zoomed-in plot of the SNR around each prominent spectral line in angular separation and physical separation bins. The figures demonstrate that the line intensity is negatively correlated with the distance between the pair for the emissions. We propose the absorption features as coming from the circumgalactic gas. The emission lines are only seen when using background galaxies, rather than sky fibers, so we expect that these lines must come from emission that is superposed on the background galaxy, but that would be rejected by the sky-fiber selection.  Possible emission sources could be 
dwarf galaxies satellites, scattered light from the galactic center by the dust around the galaxy, or direct light from the extended gaseous halo. However, we are not able to infer surface brightness as a function of distance at the outskirts of the bright galaxies because these dark background target positions are not selected randomly around a sample from BGS. Rather, DESI is preferentially choosing fainter locations, such that we would de-select brighter satellites of the bright galaxy.

\begin{figure}[htb]
    \centering
    \includegraphics[width=0.98\columnwidth]{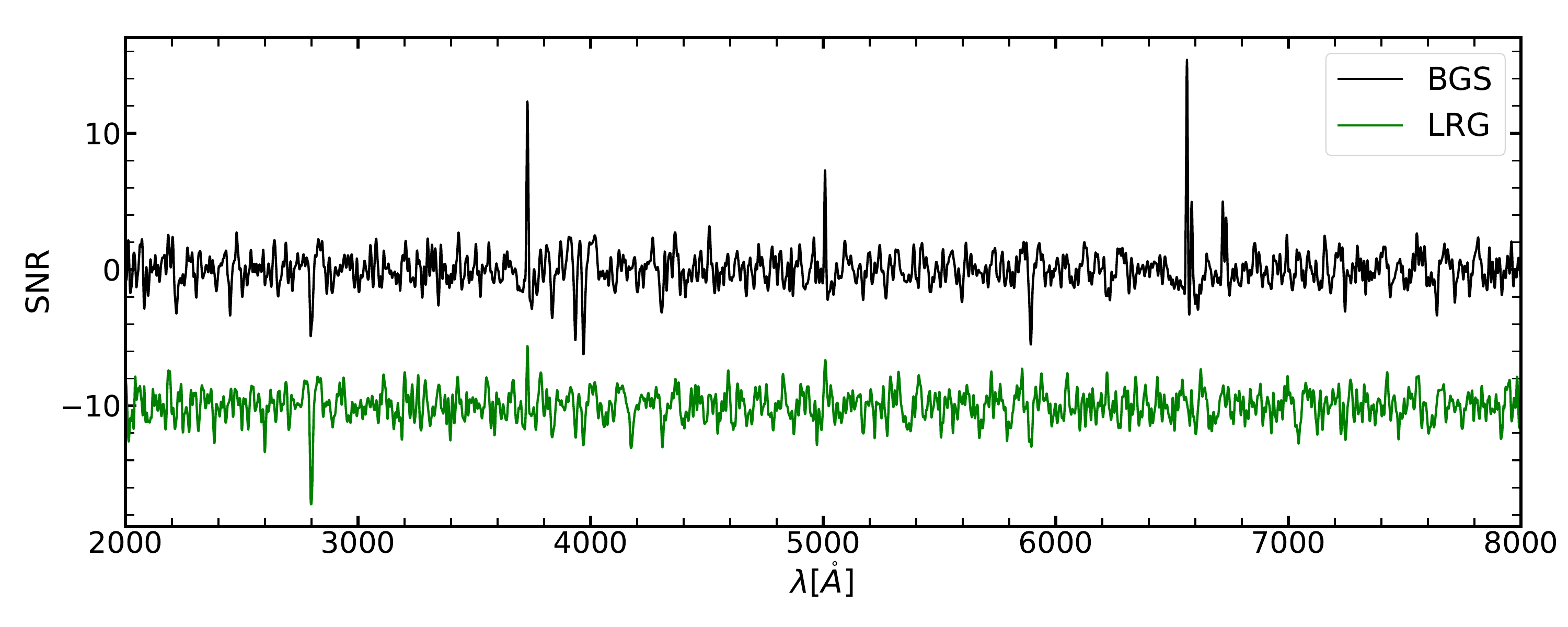}
    \caption{Comparison between stacked spectrum of BG and LRG pairs at all separations. The SNR of LRG pairs are shifted by $-10$ from $0$ to $-10$ on the plot. }
    \label{fig: com_spec}
\end{figure}

We repeat the same process for pairs with LRGs, and the comparison between stacked spectra is shown in Figure~\ref{fig: com_spec}. In the enlarged wavelength range, the $2800\,\mathring{\rm A}$ magnesium absorption line from the circumgalactic medium is also visible. We see stronger magnesium absorption features compared with the spectrum of bright galaxy pairs, which makes sense as LRGs tend to reside in denser regions with more absorbing gases around. The emission features are less prominent because the dust scattered light from dwarf galaxy neighbors is the major source for emissions, but the interstellar mediums are hot around these massive galaxies and they do not have star-forming neighbors with strong emission features. Again, we are not able to interpret the difference in line intensities between the two groups directly due to the biased target selection algorithm.


\begin{figure}[htb]
    \centering
    \includegraphics[width=0.98\columnwidth]{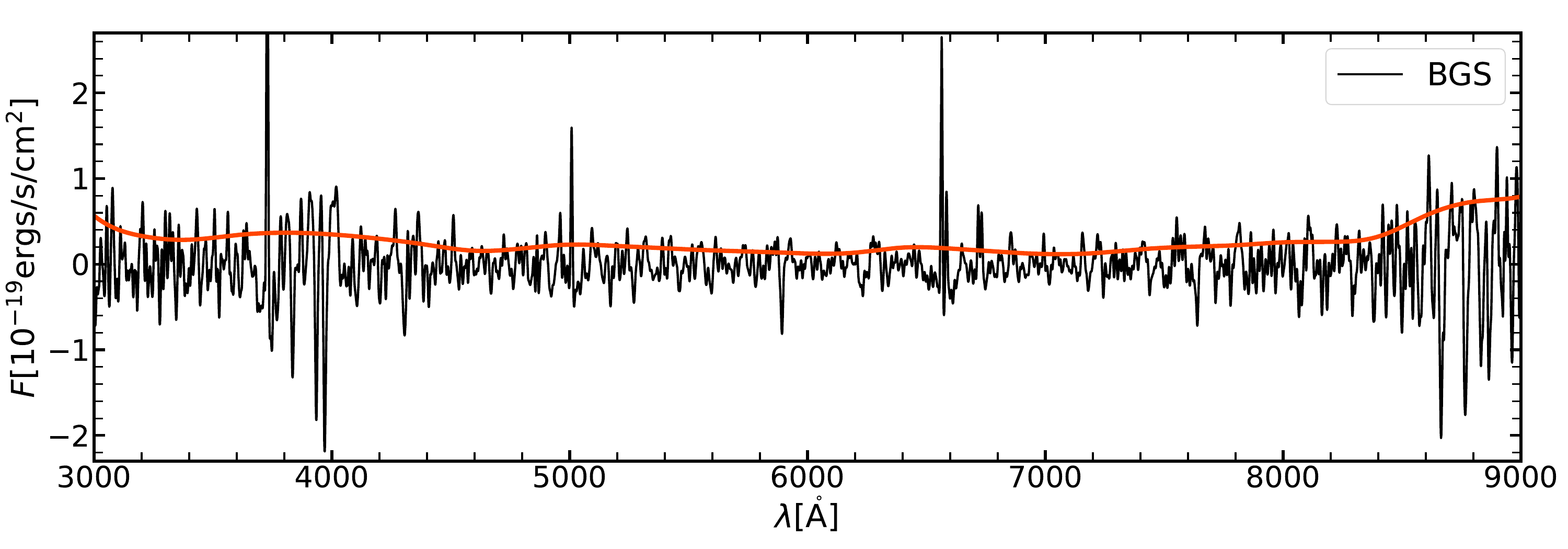}
    \includegraphics[width=0.98\columnwidth]{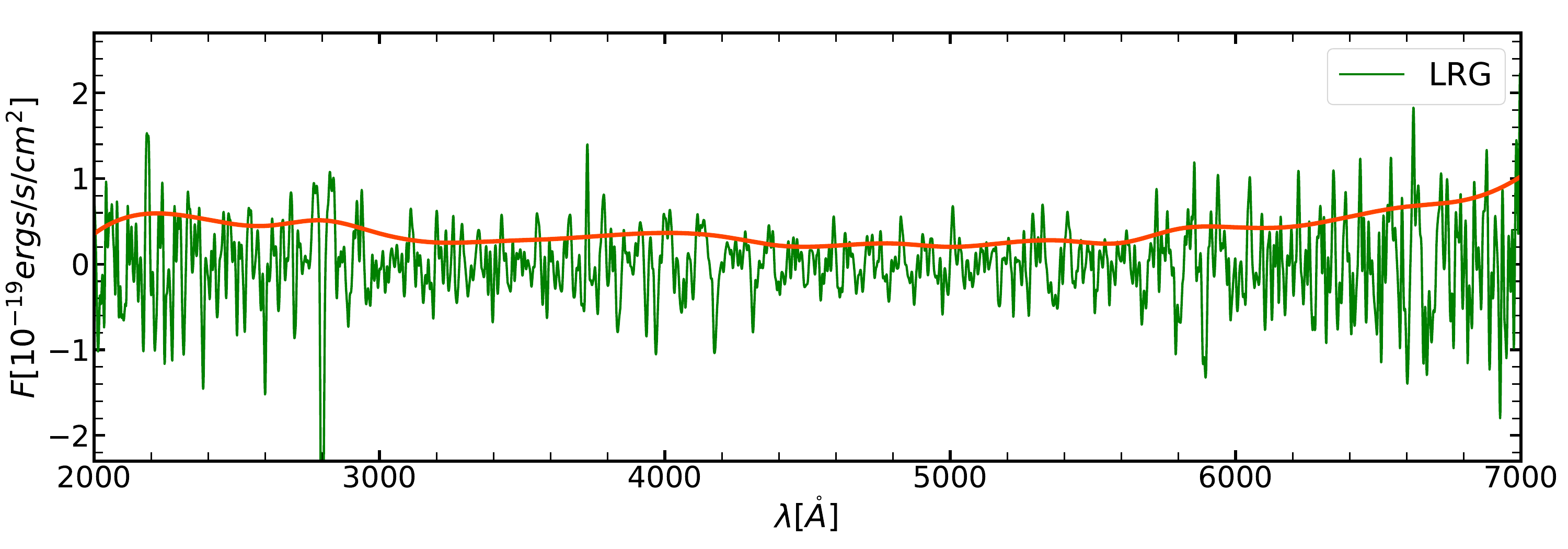}
    \caption{The convolution spectra of all blank sky and faint background objects with foreground galaxies after coaddition. The upper panels show the BG pairs and the lower panel shows the LRG pairs. The red line shows the level of noise estimated by $68\%$ error. }
    \label{fig:err}
\end{figure}

Instead, we are more interested in the general noise level of the flux density, which can be an indicator of our detection limit. 
We show in Figure~\ref{fig:err} the overall level of error in the stacked spectrum from all BG and LRG pairs including those of sky objects and of dark background targets. The result indicates that the uncertainty of flux is within $\pm 10^{-19}\,\rm ergs/\rm s/\rm cm^{2}$ for both groups.

Thus, the corresponding flux density in normal broadband centered at $5000\,\mathring{\rm A}$ is 
\begin{align}
    F_{\nu} &= \frac{10^{-19}\,\rm ergs/\rm s/\rm cm^{2}}{1.5\times10^{14}\,\rm Hz} \\
    &\sim 10^{-33}\,\rm ergs/\rm s/\rm cm^{2}/\rm Hz=10^{-10}\,\rm Jy,
\end{align}
which is $33.9$ in AB magnitude. The threshold magnitude for DESI BGS is $20$ \citep{BGS}, and the faintest object observed in Hubble eXtreme Deep Field is at AB magnitude $31$ \citep{HST}. Compared with these, our detection is much fainter than observable light from direct imaging. 

\begin{figure}[htb]
    \centering
    \includegraphics[width=0.98\columnwidth]{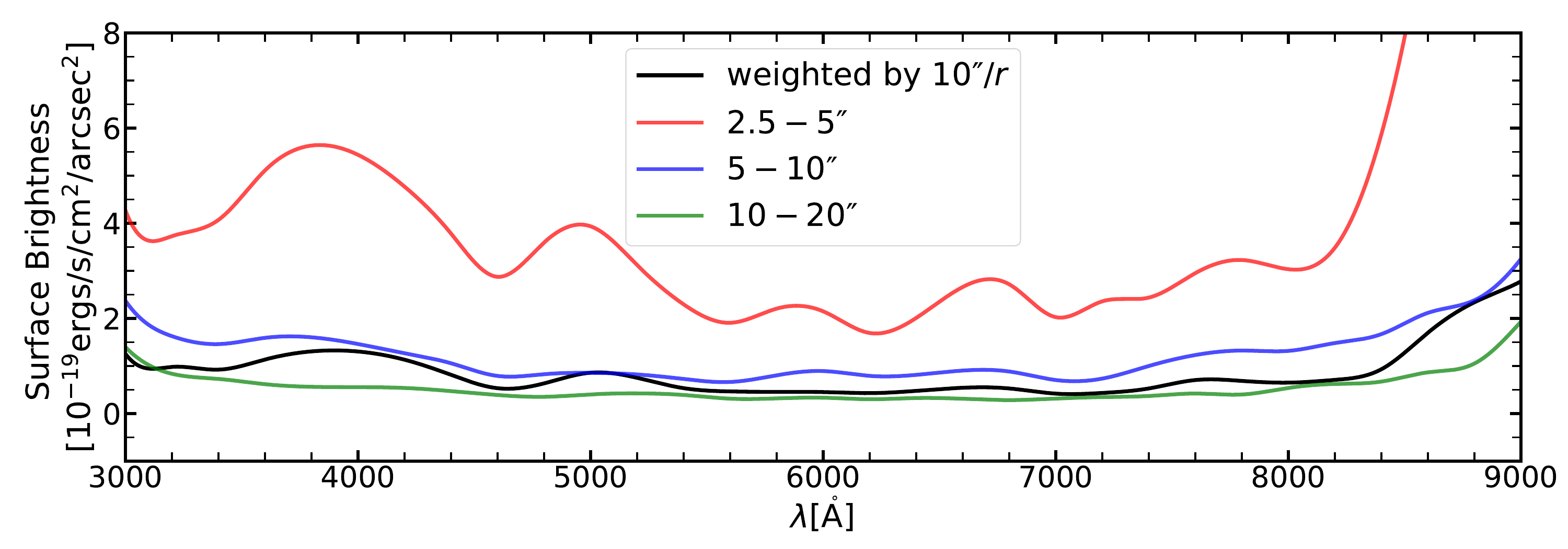}
    \includegraphics[width=0.98\columnwidth]{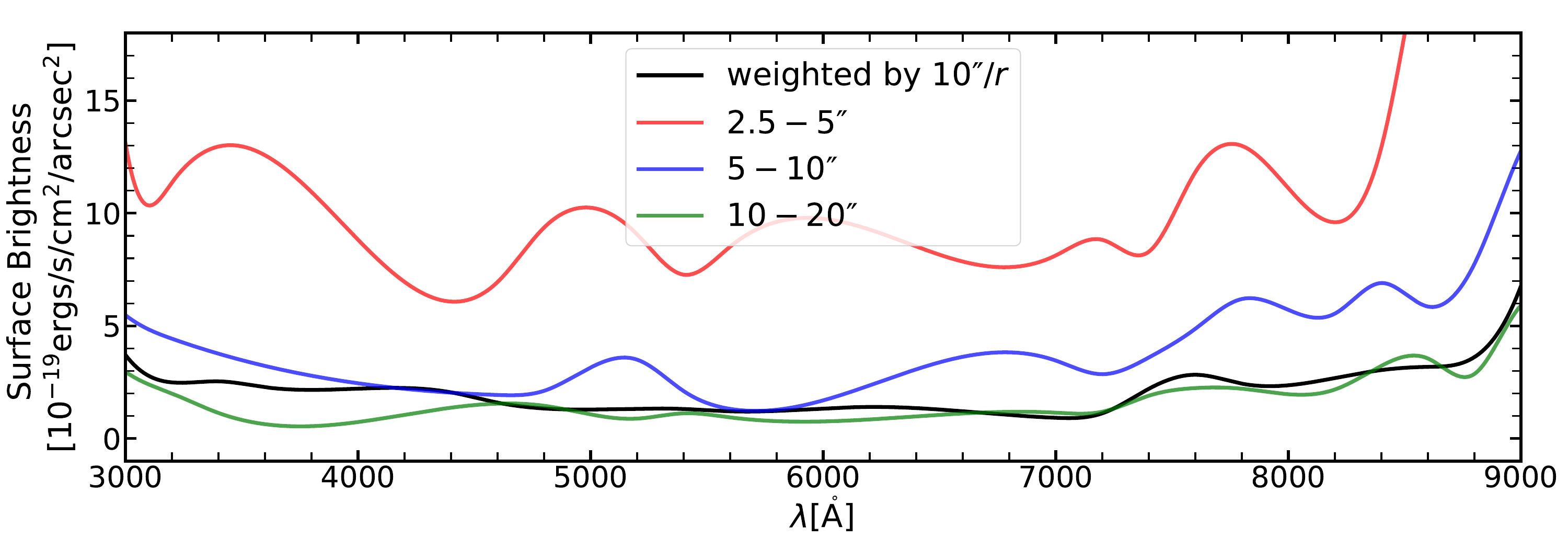}
    \caption{The $5\sigma$ detection limit of surface brightness for narrow-line emissions (top) and for broad-line emissions (bottom) generally and in angular separation bins for all BG pairs. The $5\sigma$ detection limit for narrow-line emissions is on the level of $10^{-19}\,\rm ergs/s/cm^{2}/arcsec^{2}$, whereas the detection limit for broad lines is around twice this value. The general line is calculated from the sum of all spectra weighted by $10''/r$, which is the expected model for signal strength from dark matter decay. If re-scaling the detection limits in different angular bins with this model, they appear on similar levels, while the error in the 2.5--5$''$ separation bin is a little overestimated as there are a more limited number of pairs in this subset.}
    \label{fig:5sig}
\end{figure}

With an angular diameter of $1.5''$ of the DESI fiber, the line surface brightness (flux per unit solid angle) is around the level of $10^{-19}\,\rm ergs/\rm s/\rm cm^{2}/\rm arcsec^{2}$. Moreover, fluxes should be divided by $1.3$ to undo the aperture corrections that were applied for point-source spectrophotometry.  After this correction, the $5\sigma$ detection limit for BG pairs in angular separation bins is displayed in the upper panel of Figure~\ref{fig:5sig}. We quote our results conservatively using a $5\sigma$ limit as there are many spectral channels that are being searched, such that a single-detection of 3 or even 4$\sigma$ would not be sufficient to claim a line detection.
For reference, the strongest detection of abnormal lines from the LRG sample is of $3.2\sigma$ at $2187\,\rm\mathring{A}$.

To provide a single result averaged over angle, we construct another measurement weighting each pair by $10''/r$, following the motivation that a singular isothermal sphere (SIS) density profile of $1/r^2$ would have this projected profile.
The resulting $5\sigma$ detection limit of narrow line emission at this $10''$ reference separation is about $10^{-19}\,\rm ergs/\rm s/\rm cm^{2}/\rm arcsec^{2}$, with wavelength dependence shown in the upper panel of Figure~\ref{fig:5sig}.

\begin{figure}[htb]
    \centering
    \includegraphics[width=0.98\columnwidth]{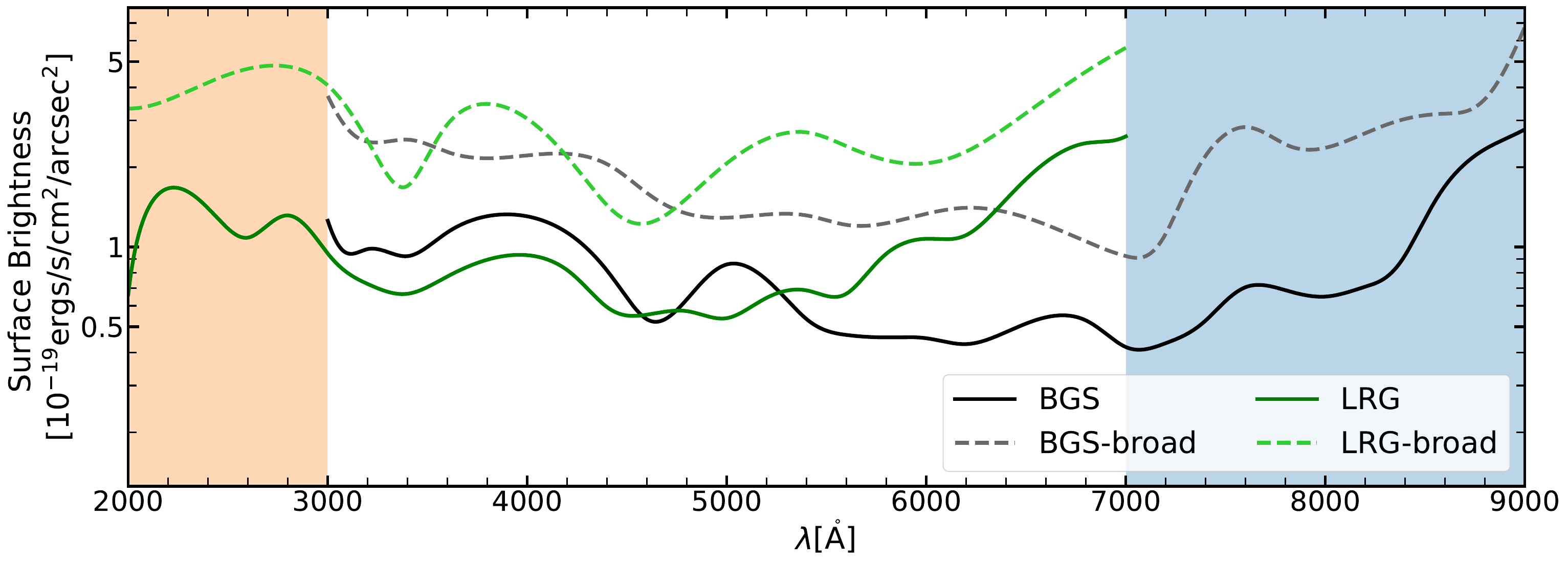}
    \caption{Comparison between the $5\sigma$ detection limit of surface brightness for narrow or broad line searches for both BG and LRG pairs. The y-axis represents values of surface brightness on a logarithmic scale,} and the shaded regions show the cutoff in the selected wavelength for each group.
    \label{fig:err_compare}
\end{figure}

While the search is mainly designed for narrow-line emissions and chooses the primary kernel size according to the dispersal velocity of galaxies, we also estimate the detection limit for broader-line emissions by increasing the sizes for both the Savitzky–Golay filter and the Gaussian convolution kernel. In the bottom panel of Figure~\ref{fig:5sig}, we show the detection limit when using a Savitzky–Golay filter with window size $505\,\mathring{\rm A}$ to remove the flux continuum and then a Gaussian convolution kernel with a width of $15\,\mathring{\rm A}$ for the same sample of BG pairs. We observe that the $5\sigma$ detection limits for broad-line emissions are less constrained than for narrow-line emissions, due to the larger size of the convolution kernel.
These results for BG pairs are again shown in Fig.~\ref{fig:err_compare}, compared with the $5\sigma$ detection limits for LRG pairs.

We repeat the same process for the LRG pairs. The weighted $5\sigma$ detection limit for narrow-line emissions and broad-line emissions is shown in Figure~\ref{fig:err_compare}, along with the comparison with those from the BG pairs.
The constraints overlap in the wavelength range 3000--7000\,\AA\ with similar emission line sensitivity but different corresponding dark matter density. 
We highlight our $5\sigma$ detection limit of surface brightness for narrow-line emissions from dark matter decay to be less than $10^{-19}\,\rm ergs/s/cm^{2}/arcsec^{2}$ for both the BG and LRG group in the wavelength range of concern, with a minimum value of approximately $0.5\times10^{-19}\,\rm ergs/s/cm^{2}/arcsec^{2}$. 
For simplicity, we cite the detection limit as $10^{-19}\,\rm ergs/s/cm^{2}/arcsec^{2}$ for future calculations.

\section{Computation of dark matter emissivity} \label{sec: comp}

The motivation of our search is to look for unidentified lines corresponding with dark matter decay. 
To interpret our detection limits in terms of a luminosity-to-mass ratio for dark matter particles, we will need to know the total mass of dark matter contributing to the measured surface brightness. 

To estimate surface mass density associated with our line search, we could adapt measurements from galaxy-galaxy weak gravitational lensing \citep{lensing}. As there are no measurements yet available with DESI BG or LRG samples, we use prior lensing signals detected with a similar flux-limited sample of low-redshift galaxy targets from the Sloan Digital Sky Survey (SDSS) for the BGS galaxies and use the lensing signals detected with the BOSS CMASS sample, a class of high-redshift color-magnitude selected galaxies, for the LRGs. While our detection limits of surface brightness are similar for BGs and LRGs, the surface dark matter densities for the two groups of galaxies are different. Therefore, we calculate dark matter emissivity based on the two samples separately.

From the Galaxy-Mass Correlation Function measured from weak lensing of galaxies in the main sample of SDSS \citep{sdss_lensing}, we assume the proper surface mass density for the BGs at $10''$ scale to be $\Sigma(R)\sim 70\,h M_{\odot}/\rm pc^{2}\sim 0.011\,\rm g/\rm cm^{2}$ for $R=40 \,\rm kpc$. Further, we assume the LRG surface mass density at $10''$ scale to be $\Sigma(R)\sim 145\,hM_{\odot}/\rm pc^{2}\sim 0.021\,\rm g/\rm cm^{2}$ for $R=70 \,\rm kpc$ in proper units, following the lensing signal measured by \cite{LRG_lensing}. 
We will compute the dark matter decay rates using these assumed values, noting that the results will scale inversely with the true values, to be measured in future DESI work.


If the dark matter had a line luminosity per unit mass of $\epsilon_{\rm DM}$, then this mass overdensity around the galaxies would produce a line flux in our apertures of solid angle $\Omega$ of 
\begin{equation}
    F = \frac{\Omega \epsilon_{\rm DM} \Sigma(R)}{4\pi (1+z)^4}
\end{equation}
where the $(1+z)^4$ arises from cosmological surface brightness dimming.
Inserting the proper surface mass densities for both galaxy samples and using the corresponding redshift ($0.24$ for BGs and $0.71$ for LRGs, obtained from the average redshift of foreground galaxies in the galaxy-spectrum pairs), we find that our $5\sigma$ surface brightness limit of $10^{-19}\,\rm ergs/s/cm^{2}/arcsec^{2}$ corresponds to a limit on the dark matter emissivity of $\epsilon_{\rm DM} \sim 10^{-5}\, \rm ergs/\rm s/\rm g$ from the BGs and $\epsilon_{\rm DM} \sim 2 \times 10^{-5}\, \rm ergs/\rm s/\rm g$ for the LRGs.

A well-motivated way to produce a monochromatic photon line, given the conservation of energy and momentum, is a two-photon decay in which the dark matter particle has a mass equal to twice the photon energy. Therefore, the wavelength range of our search from $2000\,\rm \mathring{A}$ to $9000\,\rm \mathring{A}$ corresponds to a dark matter particle mass range of $2.8$--$12.4\,\rm eV$, and the emissivity $\epsilon_{\rm DM}$ relates to the decay time $\Gamma$ by $\Gamma = \epsilon_{\rm DM}/c^{2}$. In particle physics, axion-like particles are a class of dark matter candidates in some extensions to the Standard Model \citep{ALP}. In this case, unstable axion-like particles decay into two photons with a coupling parameter $g$, yielding a decay rate 
\begin{equation}
    \Gamma= \frac{g^{2}m^{3}}{64\pi}
\end{equation}
in natural units, where $m$ is the axion-like particle mass \citep{ALP_theory}.


\begin{figure}[htb]
    \centering
    \includegraphics[width=0.98\columnwidth]{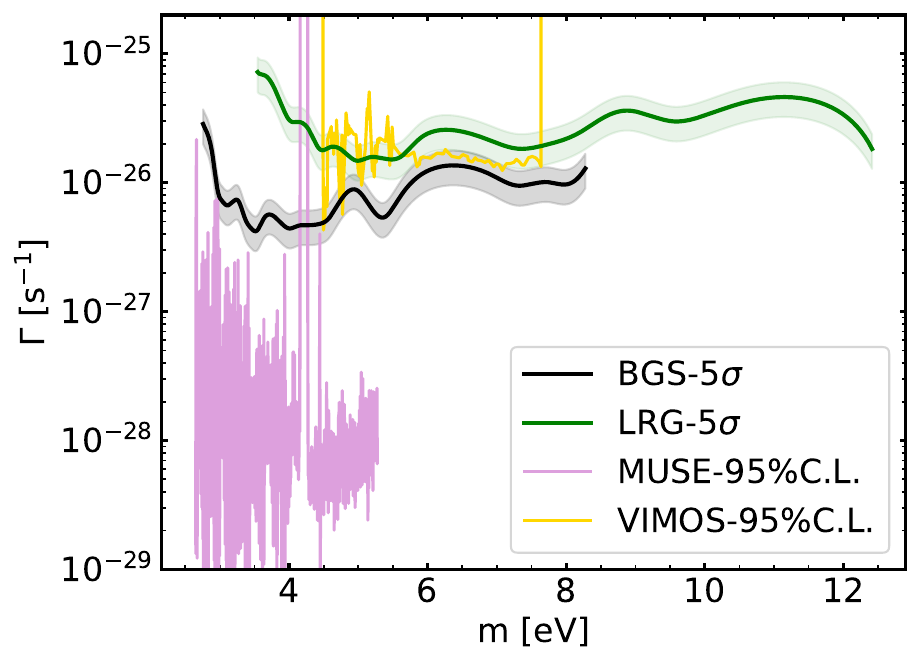}
    \caption{Bounds on the two-photon decay rate of axion-like dark matter particle derived from this work, compared with previous constraints from telescope searches. The bounds from our work are obtained from the $5\sigma$ detection limits on the surface brightness, with the shaded bands indicating the uncertainty in the surface mass density assumptions.
    The MUSE curve represents the $95\%$ one-sided confidence limit constraint from dwarf galaxy observations, and the VIMOS result is the $95\%$ limit from galaxy clusters. This difference in the confidence level produces a factor of three shift in the curves, with our result being more conservative. Nevertheless, the MUSE results are up to a factor of 30 tighter at low masses, while our results are the strongest above $5\,\rm eV$.
    }
\label{fig:gamma_compare}
\end{figure}

This relation enables us to compare our detection limit with previous bounds on the coupling of axion-like dark matter particles from telescope searches. 
In Fig.~\ref{fig:gamma_compare}, we compare the limit on the decay rate $\Gamma$ obtained from our $5\sigma$ detection limits based on the BG and LRG group with previous $95\%$ C.L. bounds measured with MUSE and VIMOS \citep{MUSE1, MUSE2, VIMOS}. 
The shaded regions around our limits indicate the systematic uncertainty resulting from the extrapolation of the surface mass density from the SDSS samples; these regions mark a 30\% variation between the yet unknown DESI surface density measurements and the assumed values based on SDSS measurements. With proper measurements of weak lensing signals around DESI BGs and LRGs in the near future, this uncertainty would be eliminated, and one could rescale the measured limits to the proper level. 
While the MUSE observations of dwarf galaxies provides tighter constraints on the particle lifetime, the MUSE instrument only covers the red optical range, restricting the particle mass range of the detection to be less than $5\,\rm eV$. 
Our search, extending to much larger particle mass using observations at higher redshifts, offers the tightest bounds on the dark matter particle lifetime in the blue and UV regime, over an order of magnitude better than the bounds probed with HST measurements of the cosmic optical background (COB) anisotropies \citep{HST_axion}.

In the particle mass range of $m \lesssim 5\,\rm eV$, the $95\%$ C.L. constraints from MUSE is approximately $30$ times better than our bounds based on the $5\sigma$ detection limit on surface brightness. We believe that the raw spectroscopic sensitivity in our DESI stack is mildly better, given the large number of galaxy-spectrum pairs.  However, the dark matter constraining power is lessened, primarily by the higher surface mass density of dark matter in the centers of dwarf galaxies compared to the density in the outskirts of more massive galaxies. Further effects are the $(1+z)^{-4}$ diminution of cosmological surface mass density in our higher redshift sample, as well the fact that the spectral lines from dwarf galaxies are narrower, allowing MUSE to take more advantage of spectral resolution in measuring the line flux.
Nevertheless, our search can still add value compared with those in dwarf galaxies, as we probe dark matter in a different physical regime, namely the outskirts of massive galaxies, where the amount of dark matter is measured accurately from weak lensing. Further, DESI data is processed with non-local sky subtraction, avoiding self-subtraction of the dwarf galaxy profile, and we use data at different redshifts, thereby mitigating the effects of strong sky emission lines.


Aside from the specific particle model, generally, $\epsilon_{\rm DM} \sim 10^{-5}\, \rm ergs/\rm s/\rm g$
is a very small emissivity, likely undetectable in terrestrial experiments.
For instance, given the dark matter density of $\sim 10^{-25}\,\rm g/\rm cm^{3}$ in our solar system \citep{DM_Solar}, the emissivity of dark matter decay is at a level of $10^{-13}$ photons per second per cubic meter. 

Beyond nearby dwarf galaxies, would we have noticed such an emissivity in other astronomical data sets? 
We first compute the impact of this emissivity on the extragalactic background light (EBL). To do this, we integrate along the line of sight to find the contribution to luminosity from the total mass of dark matter.
Using the average dark matter density of the universe $\rho_{\rm DM}=2.1\times10^{-30}\,\rm g/\rm cm^{3}$ and the dark matter emissivity derived with BG pairings, the general surface brightness after integration is 
\begin{align}
    \text{SB} &= \int_{z_{1}}^{z_{2}}\epsilon_{\rm DM}\frac{c}{H(z)}\frac{D_{A}^{2}(1+z)^{2}}{4\pi D_{L}^{2}}\rho_{\rm DM}dz \\
    &\sim \frac{c}{4\pi H_{0}}\epsilon_{\rm DM}\rho_{\rm DM}\sim 5 \times 10^{-19}\,\rm ergs/\rm s/\rm cm^{2}/\rm arcsec^{2}.
\end{align}
The EBL measured by total optical light subtracting zodiacal light (ZL) and diffuse Galactic light (DGL) is on the order of magnitude of $10^{-9}\,\rm ergs/\rm s/\rm cm^{2}/\rm sr/\mathring{A}$ in a $4000$--$7000\,\mathring{\rm A}$ filter \citep{Bernstein_2007}, which corresponds to a surface brightness around $10^{-16}\,\rm ergs/\rm s/\rm cm^{2}/\rm arcsec^{2}$.  This shows that our detection limit is more than two orders of magnitude fainter than the EBL. 

NASA’s New Horizons spacecraft has measured an anomalous component of flux coming from unknown origins at the level of $3\times10^{-17}\, \rm ergs/\rm s/\rm cm^{2}/\rm arcsec^{2}$ in the cosmic optical background (COB) using images of New Horizons’ Long Range Reconnaissance Imager (LORRI) \citep{excess-flux1,excess-flux2}, which they find is not significantly different from zero due to instrumentation uncertainties. Nevertheless, one hypothetical source of this extra flux is dark matter decay, and Line Intensity Mapping (LIM) is suggested as a useful tool for detection \citep{LIM2, LIM, Bernal}. However, the integrated surface brightness calculated with our detection limit from the co-added spectra of over three hundred thousand pairs is two orders of magnitude lower than the observed level of the anomalous flux, which makes this hypothesis implausible. 
The wavelength range from $2000$ to $9000\,\mathring{\rm A}$ in our search allows for the alternative possibility that the extra light in the COB could also be coming from an ultraviolet line at higher redshifts. It is possible to enlarge the wavelength range of the search by cross-correlating with objects at higher redshifts, such as emission line galaxies (ELGs). Nevertheless, we suggest that narrow optical-line emissions from dark matter decay cannot be the major source of the EBL.

We next consider the signal that our own Milky Way galaxy would produce from such an emissivity.  For this, we consider the surface brightness integrating along of line of sight perpendicular to the galactic plane.  As most of the signal will be generated near the solar circle, where the Milky Way has an approximately flat rotation curve, we approximate the galactic density profile as an SIS. The integration yields a surface brightness of


\begin{align}
    \rm SB &= \int \epsilon_{\rm DM} \frac{\rho_{\rm DM, solar}\cdot (8~\rm kpc/r)^{2}}{4\pi z^{2}}z^{2} dz \\
    &\sim 7\times10^{-20}\,\rm ergs/\rm s/\rm cm^{2}/\rm arcsec^{2}, 
\end{align}
where $z$ is the axis perpendicular to the galactic plane, and $r^{2}=z^{2}+(8\,\rm kpc)^{2}$. 

This signal is mildly larger than the limits we have found from the distant galaxies.  However, it is not observable in the DESI experiment because its uniformity on the degree-scale field of view of the instrument causes it to be nulled by the algorithm that measures and subtracts the foreground emission from the atmosphere and zodiacal light (known as the ``sky'' emission).

Further, this signal is so faint that it is unlikely to be recognized as not being due to the foreground emission.  To see this, we compare to the flux density of the sky emission in the optical.  Taking the predicted line surface brightness of $7\times10^{-20}\,\rm ergs/\rm s/\rm cm^{2}/\rm arcsec^{2}$ and considering the the $3\,\mathring{\rm A}$ Gaussian convolution kernel, the peak flux density of the
line surface brightness is estimated as
\begin{equation}
    \rm SB_{\lambda} \sim 10^{-20} \,\rm ergs/\rm s/\rm cm^{2}/\rm arcsec^{2}/\mathring{A}. 
\end{equation}
The optical dark-sky background is normally at AB magnitude of $22$ per square arcsecond, which corresponds to a flux density per frequency of about $10^{-5}\,\rm Jy/arcsec^{2}$ and a flux density per wavelength of
\begin{equation}
    F_{\lambda, \rm sky} = \frac{\nu}{\lambda}F_{\nu, \rm sky} \sim 10^{-17}\,\rm ergs/s/cm^{2}/arcsec^{2}/\mathring{A}.
\end{equation}

Hence, the potential signal level is three orders of magnitude lower than the sky background. With such low signal strength, given a blank sky spectrum, it can be hard to distinguish whether the anomalous contribution is from atmosphere, back scattered light from the sun, scattered light from galactic stars and interstellar medium, or actually dark matter decay in the halo. More important, practical observations subtract the sky differentially in angle, making the signal hard to detect. 
Of course, pointing toward the inner Milky Way would encounter more dark matter, but at the cost of more light from the stars and interstellar medium and more extinction from interstellar dust.  Still, we suggest that slit spectroscopic observations between the stars and utilizing the stability of a space telescope to compare background signals from tens of degrees apart might allow one to isolate the blank-field infrared emissivity of the inner galaxy. Such emissivity would almost certainly be dominated by the glow and scattering of the interstellar medium, so one would then need to look for unrecognized lines.

\section{Conclusions} \label{sec:dis}

In this work, we use positions of galaxies/dark matter to cross-correlate the spectral map of sky and faint background object emissions with DESI Y1 data. The spatial cross-correlation function gives no evidence of unidentified spectral lines caused by dark matter decay. Still, we obtain a $5\sigma$ detection limit for any potential narrow optical-line emission from dark matter of less than $10^{-19}\,\rm ergs/\rm s/\rm cm^{2}/\rm arcsec^{2}$ in the wavelength range from $2000\,$\AA\ to $9000\,$\AA. This corresponds with $34$ in normal broadband in the AB magnitude system. The mass-luminosity relation estimated from the surface mass density of galaxies indicates that dark matter decay can produce at most $3\times10^{-6}$ photon per year in a cubic meter on Earth. It is also suggested that narrow optical-line emissions from dark matter decay cannot be the major source of the EBL, as the total surface brightness contributed by all dark matter integrated along the line of sight is orders of magnitude fainter than the observed cosmic background.
If adopting the well-motivated particle model of axion-like dark matter decaying into two photons, our $5\sigma$ detection limit offers the tightest constraints on the particle lifetime in blue optical and UV regime, namely over the mass range of $5\lesssim m \lesssim 12\,\rm eV$.

Increasing observational efforts nowadays have great potential to advance astrophysical searches for dark matter. For nearby dwarf galaxies, studies of individual objects with integral field spectrographs has already proved effective. While current MUSE observations of dwarf galaxies only covers the red optical regime, an upcoming blue-optimized extension, BlueMUSE, will extend the wavelength range down to $3500\,\rm\AA$ \citep{MUSEBlue}. Meanwhile, the Keck Cosmic Web Imager (KCWI) conducts integral field spectroscopy of low surface brightness phenomena down to $3500\,\rm\AA$ \citep{KCWI}. From the same particle model, this wavelength limit of $3500\,\rm\AA$ corresponds to a particle mass of $m\sim 7\,\rm eV$ following the two-photon decay model. If with similar sensitivity as previous MUSE detection based on dwarf galaxies, these extensions can improve the current constraints on particle lifetime up to an order of magnitude for the particle mass range of $5 \lesssim m\lesssim 7\,\rm eV$.

We expect that stacking of wide-field surveys will also see substantial near-term improvements. Given the heavy investment in ground-based optical spectroscopy, stacking spectra at higher redshift is likely to be the best way to deepen the search in the rest-frame ultraviolet and hence larger particle masses. 
Wide-field surveys like DESI are expanding rapidly, and the number of spectrum-galaxy pairs available will grow as the square of the survey areal density, providing much more data that could enhance the search. Moreover, one could target regions with more dark matter, particularly using spectra of galaxies behind galaxy clusters, to provide a significant improvement in the signal strength. Eventually one must separate the dark matter signal from the diffuse starlight around the galaxies or clusters, since that will stack in the rest-frame of the foreground object and produce narrow spectral features.  However, these stellar populations are typically old and hence weak in the rest-frame UV.  We therefore expect that future analyses with DESI and newer instruments will produce substantial improvements.

\section*{Acknowledgement}
We thank the anonymous referee for the valuable comments and suggestions, which greatly improved the quality of this manuscript.
We thank Tracy Slatyer and Doug Finkbeiner for useful conversations.
H.W. was supported by the Harvard College Research Program.
D.J.E. was supported by U.S. Department of Energy grant DE-SC0007881 and as a Simons Foundation Investigator. 
This research used resources of the National Energy Research Scientific Computing Center (NERSC), a U.S. Department of Energy Office of Science User Facility located at Lawrence Berkeley National Laboratory, operated under Contract No. DE-AC02-05CH11231. 

This material is based upon work supported by the U.S. Department of Energy (DOE), Office of Science, Office of High-Energy Physics, under Contract No. DE–AC02–05CH11231, and by the National Energy Research Scientific Computing Center, a DOE Office of Science User Facility under the same contract. Additional support for DESI was provided by the U.S. National Science Foundation (NSF), Division of Astronomical Sciences under Contract No. AST-0950945 to the NSF’s National Optical-Infrared Astronomy Research Laboratory; the Science and Technology Facilities Council of the United Kingdom; the Gordon and Betty Moore Foundation; the Heising-Simons Foundation; the French Alternative Energies and Atomic Energy Commission (CEA); the National Council of Science and Technology of Mexico (CONACYT); the Ministry of Science and Innovation of Spain (MICINN), and by the DESI Member Institutions: \url{https://www.desi.lbl.gov/collaborating-institutions}. Any opinions, findings, and conclusions or recommendations expressed in this material are those of the author(s) and do not necessarily reflect the views of the U. S. National Science Foundation, the U. S. Department of Energy, or any of the listed funding agencies.

The authors are honored to be permitted to conduct scientific research on Iolkam Du’ag (Kitt Peak), a mountain with particular significance to the Tohono O’odham Nation.

\section*{Data Availability}
All data points shown in the published graphs are available in machine-readable form in Zenodo at \url{https://doi.org/10.5281/zenodo.8339781}.

\bibliography{main.bib}

\end{document}